\documentclass[12pt]{article}

\usepackage[a4paper,text={16.0cm,23.0cm},centering]{geometry}
\usepackage{latexsym}
\usepackage{graphicx}
\usepackage{amsmath}
\usepackage{amssymb}
\usepackage{mathrsfs}
\usepackage{bm}
\usepackage{fancybox}
\usepackage{color}
\usepackage{wrapfig}
\usepackage{framed}
\usepackage{mathtools}

\def\rd{{\rm d}}
\def\fr{\mbox{$\frac{1}{2}$}}
\def\R{\mathbb R}
\def\qand{\quad\mbox{and}\quad}
\def\Rg{{\mathcal R}}
\def\bO{{\bm\Omega}}

\begin{document}

\begin{center}
  \textsf{\Large A variational principle for fluid sloshing}\\[2mm]
  \textsf{\Large with vorticity, dynamically coupled to vessel motion}
\vspace{.5cm}

\textit{H. Alemi Ardakani$^1$, T.J. Bridges$^2$, F. Gay-Balmaz$^3$, Y. Huang$^2$
, \& C. Tronci$^2$}
\vspace{.25cm}

\textit{\phantom{*}$^1$ Department of Mathematics, University of Exeter,
  Penryn Campus}\\
\textit{Penryn, Cornwall TR10 9FE, UK}\\
\textit{\phantom{*}$^2$Department of Mathematics, University of Surrey, Guildfor
d GU2 7XH, UK}\\
\textit{\phantom{*}$^3$Laboratoire de M\'{e}t\'{e}orologie Dynamique,
  \'{E}cole Normale Sup\'{e}rieure \& CNRS},\\
\textit{F-75231 Paris, France}
\vspace{1.0cm}

\setlength{\fboxsep}{10pt}
\doublebox{\parbox{12cm}{
{\bf Abstract.} 
A variational principle is derived for two-dimensional incompressible
rotational fluid flow with a
free surface in a moving vessel when both the vessel and
fluid motion are to be determined.
The fluid is represented by a stream function
and the vessel motion is represented by a path in the planar Euclidean group.
Novelties in the formulation include how the pressure
boundary condition is treated, the introduction of a stream function into
the Euler-Poincar\'e variations, the derivation of free surface variations,
and how the equations for the vessel path in the Euclidean group, coupled
to the fluid motion, are generated automatically.  
}}
\end{center}
\hfill{\textsf{\today}}

\section{Introduction}
\setcounter{equation}{0}
\label{sec-intro}

Variational principles for fluid sloshing; that is, free surface flow in
an enclosed container, abound, having first been derived 
independently by Lukovsky \cite{lukovsky-76} and Miles \cite{miles}, based on Luke's
variational principle \cite{luke}.  Variational principles have since been
widely used in the analysis and modelling of fluid sloshing (e.g.\ Chapters 2,7 of \cite{slosh-book} and references therein).  However, in all this work
the fluid motion is assumed to be irrotational.  A variational principle
for fluid sloshing with vorticity was first introduced by \textsc{Timokha}~\cite{timokha} using Clebsch variables to represent the velocity field. In two
space dimensions (2D), Clebsch
variables represent the vorticity field exactly, but they are difficult
to work with and have singularities in the potentials even when the velocity
field is smooth (e.g.\  \textsc{Fukagawa \& Fujitani}~\cite{ff10},
\textsc{Graham \& Henyey}~\cite{gh00}).

In this paper a new variational principle for fluid flow with
a free surface and vorticity is obtained in 2D by representing the velocity field
in terms of a stream function and using constrained variations.
Moreover, the variational principle captures dynamic coupling with
the vessel motion; that is, the variational principle produces both the
exact fluid equations in
a moving vessel, as well as coupling with
the exact equations of motion for the vessel.  The
vessel motion is a path in the special Euclidean group, $SE(2)$,
consisting of rotations and translation in the plane.

Variational principles for the coupled motion have been given before, the first by
\textsc{Lukovsky} (see \cite{lukovsky} and references therein), for
the fluid-vessel coupling, and more recently
by \textsc{Alemi Ardakani}~\cite{haa-thesis,haa17}, which, in the latter
case, includes
coupling between the vessel and both interior and exterior fluid
motion.  However, in all these
cases the fluid motion is taken to be irrotational.
Here the aim is to dynamically couple
vessel motion to interior fluid flow with an exact and complete
representation of the vorticity field.

Variational principles for Eulerian fluid motion with vorticity
are notoriously difficult.
On the other hand, variational principles in the Lagrangian particle path (LPP)
formulation are the natural continuum versions of Hamilton's principle in classical mechanics and are therefore relatively straightforward.  For example, \textsc{Hsieh}~\cite{hsieh}
presents a variational principle for the LPP formulation
with a free surface, based on the kinetic minus the potential energy,
which is used to model bubble dynamics, and no special
constraints or constrained variations are required.
However, the LPP formulation is not as
useful in practice as the Eulerian fluid representation.

Natural variational principles, starting with the kinetic minus the
potential energy, fail in the Eulerian setting.  For
example, take the 2D
fluid domain to be in the region $0<y<h(x,t)$ where $y=h(x,t)$ represents
the free surface for $0\leq x\leq L$.  Then the natural Lagrangian, based on the kinetic minus potential energy, when the velocity is represented by a stream function, is
\begin{equation}\label{K-V-Lagr}
  \delta\int_{t_1}^{t_2}\mathcal{L}(\psi,h)\,\rd t=0\quad\mbox{with}\quad
  \mathcal{L}(\psi,h) = \int_0^L \int_0^h \left[
  \fr\rho (\psi_x^2+\psi_y^2) - \rho g y \right] \rd x\rd y\,,
\end{equation}
where $\rho$ is the constant fluid density and $g$ is the gravitational constant.
Taking free variations with respect to $h$ and $\psi$ and setting
$\delta\mathcal{L}/\delta\psi$ and $\delta\mathcal{L}/\delta h$ to zero
does not lead to the correct governing equations or boundary conditions.
The reason being that $\psi$ and $h$ are not the Lagrangian variables of the problem, and hence the classical Hamilton principle with free variations does not apply when such variables are used.

One thus needs to consider the \emph{constrained variations} induced by the free variations of the Lagrangian variables. For fluids with a fixed boundary, this is known as the \emph{Euler-Poincar\'e framework} (e.g.\ Chapter 11
of \textsc{Holm et al.}~\cite{hss09}). 
In the Euler-Poincar\'e framework, the constrained variation of the velocity field is given by
\begin{equation}\label{delta-u-def-intro}
  \delta{\bf u} = \mathbf{z}_t + [{\bf u},\mathbf{z}]\,,
\end{equation}
where ${\bf z}$ is a vector-valued
free variation, and $[\cdot,\cdot]$ is the Lie bracket of vector fields \cite{hmr98,hss09} (that is, $[\mathbf{u},\mathbf{z}]=\mathbf{u}\cdot\nabla\mathbf{z}-\mathbf{z}\cdot\nabla\mathbf{u}$).
However, we will need to introduce two extensions of this theory: firstly, inclusion
of a free boundary $h(x,t)$ and an appropriate variation $\delta h$,
and secondly, how to induce a constrained variation, $\delta\psi$, for
the stream function. 

The Euler-Poincar\'e framework was first extended to free boundary flows in
\cite{gbms12} with a compressible fluid in the interior.  The
incompressible case can be obtained {\it a posteriori} by setting density to
be constant.  
Here, the new strategy is to address the incompressible case directly
by working with divergence-free vector fields
from the start, parameterized with a stream function, then use (\ref{delta-u-def-intro})
and a Lie algebra homomorphism to obtain $\delta\psi$ directly,
\begin{equation}\label{delta-psi-def-intro}
\delta\psi = w_t + \{w,\psi\}\,,
\end{equation}
where $w(x,y,t)$ is scalar-valued and a free variation,
and $\{\cdot,\cdot\}$ is the standard $(x,y)-$Poisson bracket for scalar
valued functions.  This variation, and a reduction from the LPP
setting to the Eulerian setting, induces a variation at the free surface
\[
\delta h = - W_x\,,\quad W(x,t) := w(x,h(x,t),t)\,.
\]
The expressions for $\delta\psi$ and $\delta h$ are proved in
\S\ref{sec-reduction}.  They are derived from first principles using
reduction from the LPP to Eulerian formulation.

The difficulties with the free boundary are compounded by the use of
a stream function formulation: the pressure no longer appears
explicitly rendering the dynamic
 boundary condition at the free surface problematic.  This problem is
 resolved in a novel way by showing that the pressure boundary condition
 is equivalent to the kinematic conservation law of
 \textsc{Gavrilyuk et al}~\cite{gkk15} (hereafter GKK conservation law),
 extended to the case of
 free surface flow relative to a moving frame. Then
 a key result of the variational construction is how the GKK conservation law emerges
from the variational principle, justifying this new form for the pressure
boundary condition.

The variational principle is useful for establishing structure,
identifying conservation laws, constructing numerical schemes,
and developing approximate methods such as the multimodal expansion of
solutions.  

An outline of the paper is as follows.  Firstly, the governing equations
for the coupled problem are written down in \S\ref{sec-goveqns}.
Then a Lagrangian density is formulated based on the kinetic minus potential
energy of the fluid and vessel motion, relative to a moving frame in
\S\ref{sec-variationalformulation}.  Variations are then taken in the
directions $\delta h$, $\delta\psi$, $\delta{\bf q}$ and $\delta\Rg$,
where ${\bf q}$ is the body translation vector in 2D and
$\Rg$ is a rotation matrix in the plane representing the body orientation.
Special cases of the resulting equations are given in \S\ref{sec-specialcases}.
The justification of the expressions for the constrained variations is
given in \S\ref{sec-reduction} based on reduction from the LPP setting to the
Eulerian setting.  In the concluding remarks section \S\ref{sec-cr} some implications
and potential extensions of the new variational formulation are discussed.  

\section{Governing equations}
\setcounter{equation}{0}
\label{sec-goveqns}

The fluid is incompressible and of constant density $\rho$.
The fluid occupies the two-dimensional (2D) region
\begin{equation}\label{fluid domain}
\mathcal{D}:=
\big\{ (x,y)\in\R^2\ :\ 0<y<h(x,t)\ \mbox{and}\quad 0<x<L\big\}\,,
\end{equation}
for some $L>0$ and $y=h(x,t)$ is a graph representing the free surface and
it is to be determined.
The fluid equations are relative to the body-fixed
frame with coordinates ${\bf x}=(x,y)$ and its relation to the
spatial frame is given below in (\ref{body-space-coord}).

Upon denoting the velocity field by $\mathbf{u}=(u,v)$, the
governing equations for the velocity and pressure, relative to the
body frame, are
\begin{equation}\label{euler-rotating}
\begin{array}{rcl}
  \displaystyle \frac{Du}{Dt} + \frac{1}{\rho}\frac{\partial p}{\partial x}
  &=& -g\sin\theta +2\dot\theta v + \ddot\theta y + \dot\theta^2x
  - \ddot q_1\cos\theta - \ddot q_2 \sin\theta \\[4mm]
\displaystyle \frac{Dv}{Dt} + \frac{1}{\rho}\frac{\partial p}{\partial y}
  &=& -g\cos\theta -2\dot\theta u - \ddot\theta x + \dot\theta^2y
  + \ddot q_1\sin\theta - \ddot q_2 \cos\theta \,,
  \end{array}
\end{equation}
where $g>0$ is the gravitational constant and
$\frac{D f}{Dt}
:= \frac{\partial f}{\partial t} + u\frac{\partial f}{\partial x}
+ v\frac{\partial f}{\partial y}$.
The functions $(\theta,q_1,q_2)$ represent the orientation and translation of
the body.  These equations are derived in \cite{haa-thesis} and in
\S2 of \cite{aab12}.
Conservation of mass relative to the body frame is
\begin{equation}\label{consv-mass}
u_x + v_y = 0\,,
\end{equation}
which also acts as an equation for the pressure.
The boundary conditions on the vessel walls are
\begin{equation}\label{rwcs}
u=0\quad\mbox{at}\quad x=0,L\quad\mbox{and}\quad v=0
\quad\mbox{at}\quad y=0\,,
\end{equation}
and at the free surface the boundary conditions are
\begin{equation}\label{fsbcs}
p = 0 \qand h_t + uh_x =v \quad\mbox{at}\quad y=h(x,t)\,.
\end{equation}
The free-surface boundary condition on the pressure is obtained by assuming that surface tension effects are neglected and by requiring the pressure to be equal to atmospheric pressure at the surface. Since $p$ is defined up to
an arbitrary function of time,
we can choose $p=0$ at the surface. The free surface equation for the height directly follows from the definition of the variable $h(x,t)$ in terms of the Lagrangian fluid motion $\mathbf{x}=\varphi(t,\mathbf{a})$. This Lagrangian
point of view is developed in \S\ref{sec-reduction}.

\begin{figure}[!h]
\centering\includegraphics[width=7cm]{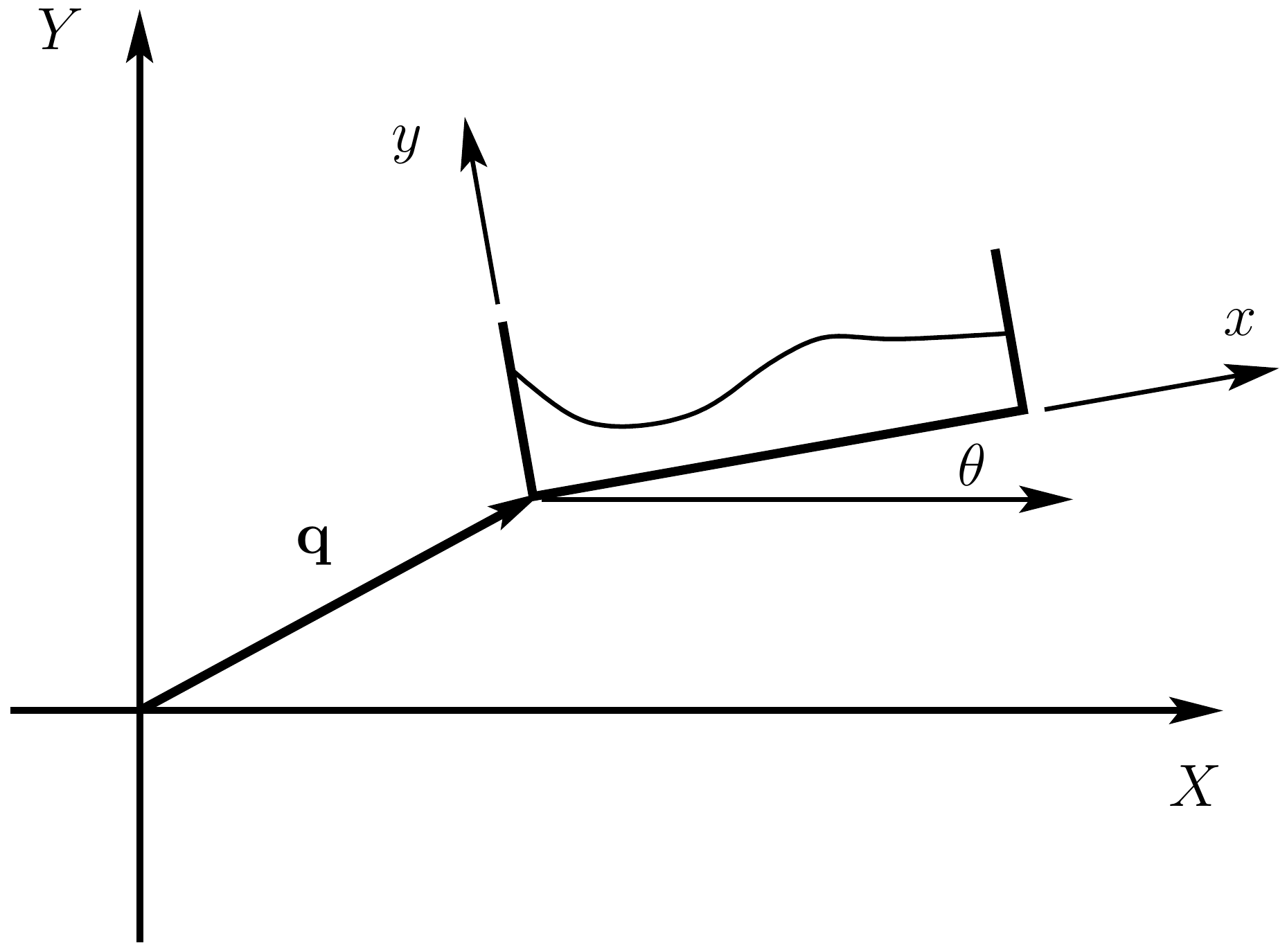}
\caption{Schematic of the fluid domain relative to the body frame.}
\label{fig-q-theta-vessel}
\end{figure}

The body-fixed frame with coordinates
${\bf x}=(x,y)$ is related to the space-fixed frame with
coordinates ${\bf X}=(X,Y)$ by
\begin{equation}\label{body-space-coord}
  {\bf X} = \Rg(t){\bf x} + {\bf q}(t)\,,
\end{equation}
where ${\bf q}=(q_1,q_2)$ represents uniform translation of the frame, and
\begin{equation}\label{Q-def}
\Rg(t) = \left[\begin{matrix} \cos\theta(t) & -\sin\theta(t) \\
      \sin\theta(t) & \cos\theta(t)\end{matrix}\right]\,.
\end{equation}
A schematic is shown in
Figure \ref{fig-q-theta-vessel}.
When the axis of rotation is at another point other than the origin
of the body axis then (\ref{body-space-coord}) is replaced by
${\bf X} = \Rg(t)\big({\bf x}+{\bf d}\big) + {\bf q}(t)$ where
${\bf d}$ is a constant vector. The theory will be developed
for ${\bf d}=0$, to simplify notation,
as the shift in axis of rotation
can be added in {\it a posteriori} as required.

\subsection{Stream function and vorticity}
      \label{subsec-streamfunction}
      
      In 2D divergence free vector fields in
      $\mathcal{D}$ can be parameterised by a stream function,
      \begin{equation}\label{psi-omega-def}
      u = \psi_y\,,\quad v = -\psi_x\,.
      \end{equation}
      Conservation of mass (\ref{consv-mass}) is satisfied exactly and the
      vorticity $\mathcal{V}$ is defined by
      \begin{equation}\label{vort-eqn}
        \mathcal{V} = v_x -u_y = -\Delta\psi\qand
        \frac{D\mathcal{V}}{Dt} = -2\ddot\theta\,.
      \end{equation}
      Substitute the stream function into the free surface boundary
      condition (\ref{fsbcs}),
      \begin{equation}\label{kfsbc-a}
      0 = h_t+uh_x-v = h_t + \psi_yh_x + \psi_x = h_t + \Psi_x \,,
      \end{equation}
      where
      \begin{equation}\label{Psi-def}
        \Psi(x,t) = \psi(x,h(x,t),t)\,,
      \end{equation}
      giving the following
      form for the kinematic free surface boundary condition
      \begin{equation}\label{kfsbc}
        h_t + \Psi_x = 0 \quad \mbox{at}\quad y=h(x,t)\,.
        \end{equation}

      At this point a difficulty emerges with the pressure boundary condition, $p=0$ at $y=h$, since pressure does not appear
          in the vorticity-stream function formulation.  However, the Euler equations have a gauge symmetry in that an arbitrary function
of time can be added to the pressure field without changing the dynamics.
Hence the more general boundary condition for the pressure is $p=0$ modulo an arbitrary function of time, or
\begin{equation}\label{pressure-bc-2}
0 = \frac{\partial\ }{\partial x} p(x,h(x,t),t) = p_x + p_yh_x \quad\mbox{at}\quad y=h(x,t)\,.
\end{equation}
Substituting for $p_x$ and $p_y$ from (\ref{euler-rotating}) then gives
\begin{equation}\label{pressure-bc}
\begin{array}{rcl}
  \displaystyle\frac{Du}{Dt} +h_x \frac{Dv}{Dt}
  &=& -g\sin\theta +2\dot\theta v + \ddot\theta y + \dot\theta^2x
  - \ddot q_1\cos\theta - \ddot q_2 \sin\theta \\[2mm]
&&\quad +h_x (-g\cos\theta -2\dot\theta u - \ddot\theta x + \dot\theta^2y
  + \ddot q_1\sin\theta - \ddot q_2 \cos\theta),
  \end{array}
\end{equation}
at $y=h(x,t)$.
Although this boundary condition looks complicated it has an elegant and simple form,
\begin{equation}\label{gkk-claw}
K_t + (uK - \Xi)_x = 0 \,,
\end{equation}
with
\begin{equation}\label{K-def-2D}
  K = \rho\big[u -\dot\theta h + v_1 + h_x(v + \dot\theta x + v_2)\big]\,,
  \end{equation}
  and flux function
  \begin{equation}\label{Xi-condensed}
  \Xi = \fr\rho(u-\dot\theta h+v_1)^2+\fr\rho(v+\dot\theta x+v_2)^2 -\rho (\Gamma_1x+\Gamma_2h)\,,
  \end{equation}
  where $\boldsymbol{\Gamma}=(\Gamma_1, \Gamma_2)$ and ${\bf v}=(v_1,v_2)$ are the gravity vector and vessel acceleration
  vector \emph{relative to the body frame} and explicit expressions for them are
  given in \S\ref{sec-variationalformulation} below.

  The
  conservation law (\ref{gkk-claw}) is a generalization of the
  GKK conservation law \cite{gkk15} to the case of free boundary
  flow relative to a moving frame.
  When $\theta=\dot\theta=v_1=v_2=0$ the above conservation law reduces to
  \begin{equation}\label{gkk-original}
      K_t + (uK + \rho gh - \fr \rho u^2 - \fr \rho v^2)_x=0\,,\quad K=\rho (u+vh_x)\,,
      \quad\mbox{at}\ y=h(x,t)\,,
  \end{equation}
      which is the form of the \emph{kinematic conservation law} given
      in equation (14) of \cite{gkk15}.

          \subsection{Summary of the fluid equations}
          \label{subsec-summary-fluideqns}
          
          In summary, in the vorticity-stream function formulation,
      the governing equations (\ref{euler-rotating}) are replaced by
      \begin{equation}\label{gov-eqns-psi}
        \mathcal{V} = -\Delta\psi\qand \frac{D\mathcal{V}}{Dt}=-2\ddot\theta\quad\mbox{in}\ \mathcal{D}\,.
      \end{equation}
The boundary conditions at the solid walls are
      \begin{equation}\label{bc-wall-psi}
        \psi_x= 0\quad\mbox{at}\ y=0\,,\quad \psi_y=0\quad\mbox{at}\ x=0,L\,.
      \end{equation}
      The boundary conditions at the free surface are
      \begin{equation}\label{kfsbc-psi}
      h_t + \Psi_x = 0\qand K_t + (\psi_yK -\Xi)_x =0 \quad\mbox{at}\ y=h(x,t)\,,
      \end{equation}
      with $K$ and $\Xi$ given in (\ref{K-def-2D})
      and (\ref{Xi-condensed}) respectively, with $u,v$ replaced by their
      stream function representation.
      The aim is the show that these governing equations for the fluid
      follow from a variational principle. 
           
      \subsection{Summary of the vessel equations}
      \label{subsec-vessel-eqns}

      The fluid equations in
      \S\ref{sec-goveqns}\ref{subsec-summary-fluideqns} are
      dynamically coupled to the vessel motion.  The vessel
      position and orientation are given in terms of
      $\theta(t)$ and ${\bf q}(t)$ and they are coupled to the fluid motion.
      Governing equations for $(\theta(t),{\bf q}(t))$ can be deduced from
      Newton's law for the total linear and angular momentum.

      The translation of the vessel follows from the conservation of
  total linear momentum
  \begin{equation}\label{claw-total-lin-mom}
           \frac{d\ }{dt}\int_0^L\!\!\int_0^h \rho \Rg\big( {\bf u} +
           \boldsymbol{\Omega}\times{\bf x} + \Rg^T\dot{\bf q}\big)\,\rd y \rd x = -\int_0^L\!\!\int_0^h \rho g{\bf E}_2\,\rd y\rd x\,.
  \end{equation}
           The orientation of the vessel follows from the conservation of total
           angular momentum
           \begin{equation}\label{claw-total-ang-mom}
           \frac{d\ }{dt}\left[\int_0^L\!\!\int_0^h
           \rho(\Rg{\bf x}+{\bf q})\times \Rg({\bf u}+\boldsymbol{\Omega}\times{\bf x}+
                   \Rg^T\dot{\bf q})\,\rd y\rd x \right] = -\int_0^L\!\!\int_0^h\rho g(\Rg{\bf x}+{\bf q})\times {\bf E}_2\,\rd y \rd x\,.
           \end{equation}
           In these equations
           \[
           \boldsymbol{\Omega}= \dot\theta{\bf e}_3
           = \dot\theta{\bf E}_3
           = \boldsymbol{\omega}\,,
           \]
           that is, the body angular velocity $\Omega$ equals the
           space angular velocity $\omega$.  The basis for the
           spatial frame is denoted by $({\bf E}_1,{\bf E}_2,{\bf E}_3)$
           and the basis for the body frame is denoted by
           $({\bf e}_1,{\bf e}_2,{\bf e}_3)$.
           The fluid velocities $\mathbf{U}$ and $\mathbf{u}$
           in the spatial and body frame, respectively,
are related by
\begin{equation}\label{U-u-def}
    {\bf U} = \dot\Rg{\bf x} + \Rg{\bf u}+\dot{\bf q}=
    \Rg\left[
      {\bf u} + {\bm\Omega}\times {\bf x} + \Rg^T\dot{\bf q}\right]\,.
\end{equation}
  The governing equations for the vessel, (\ref{claw-total-lin-mom})
  and (\ref{claw-total-ang-mom}), can be expanded and simplified, but
  it is easier to develop these equations in the form that emerges
  from the variational principle.

\section{Lagrangian of the coupled system}
\setcounter{equation}{0}
\label{sec-variationalformulation}

The proposed variational principle is
\begin{equation}\label{delta-L}
\delta\int_{t_1}^{t_2} \mathcal{L}\, \rd t = 0\,,
\end{equation}
where the Lagrangian is the kinetic minus potential energy
\begin{equation}\label{Lagr-density}
\mathcal{L} =
\textsf{KE}^f + \textsf{KE}^v - \textsf{PE}^f - \textsf{PE}^v\,,
\end{equation}
where the superscript $f$ indicates fluid and $v$ the vessel.  These
energies are formulated as follows.

  The energy densities of the fluid are
  \begin{equation}\label{KE-fluid-integral}
    \textsf{KE}^f = \int_0^L \int_0^h\widehat{\textsf{KE}^f}\,\rd y\rd x
    \qand  \textsf{PE}^f:= \int_0^L\int_0^h\widehat{\textsf{PE}^f}\,\rd y\rd x\,,
    \end{equation}
    with
  \begin{equation}\label{KE-fluid-density}
    \begin{array}{rcl}
      \widehat{\textsf{KE}^f} &=&
      \displaystyle \fr \rho \|{\bf U}\|^2 \\[2mm]
      &=& \fr\rho \| {\bf u} + {\bm\Omega}\times{\bf x}+\Rg^T\dot{\bf q}\|^2
      \quad\mbox{(using (\ref{U-u-def}))}\\[2mm]
&=& \fr\rho\left( {\bf u} + {\bm\Omega}\times{\bf x} + {\bf v}\right
)\cdot\left( {\bf u} + {\bm\Omega}\times{\bf x} + {\bf v}\right)\\[2mm]
    &=& \fr\rho\|{\bf u}\|^2 + \rho{\bf u}\cdot{\bm\Omega}\times{\bf x}
    +  \fr\rho\|{\bm\Omega}\times{\bf x}\|^2 + \rho{\bf v}\cdot{\bm\Omega}\times{\bf x} +
    \rho{\bf v}\cdot{\bf u}  + \fr\rho\|{\bf v}\|^2\,.
  \end{array}
\end{equation}
  The vessel position and velocity and the gravity vector are represented
  relative to the body frame.
  \begin{equation}\label{Gamma-r-v-def}
  \Gamma = g\Rg^T{\bf E}_2\,,\quad {\bf r}=\Rg^T{\bf q}\,,\qand
         {\bf v} = \Rg^T\dot{\bf q}\,.
  \end{equation}
  With these definitions the kinetic and potential energies are
  $\Rg-$invariant, and there is no explicit dependence on the group
  action. 

  Re-introducing integration, the fluid kinetic energy takes the form
\[
  \begin{array}{rcl}
    \textsf{KE}^f &=&\displaystyle \int_0^L\!\!\int_0^h \Big( \fr \rho \|{\bf u}
\|^2 + \rho{\bf u}\cdot{\bm\Omega}\times{\bf x} + \rho {\bf v}\cdot{\bf u}  \Big)\rd 
y \rd x
    \\[6mm]
    &&\displaystyle\hspace{2.0cm} + \fr \Pi^f{\bm\Omega} \cdot{\bm\Omega}  - m_f\overline{
\bf x}^f\cdot {\bm\Omega}\times{\bf v} + \fr m_f \|{\bf v}\|^2 \,.
  \end{array}
  \]
  where
  \[
  m_f = \int_0^L\!\!\int_0^h\rho \,\rd y \rd x= \int_0^L \rho h(x,t)\,\rd x\,,\quad
  \overline{\bf x}^f =\frac{1}{m_f}\int_0^L\!\!\int_0^h\rho {\bf x}\,\rd y \rd x
\,,
  \] 
  and
  \[  
  \Pi^f := \int_0^L\!\!\int_0^h\rho\Big( \boldsymbol{1}\|{\bf x}\|^2- {\bf x}{\bf x}^T\Big)\rd y \rd x \,.
  \]
  In the 2D case considered here the moment of inertia term reduces to
  \begin{equation}\label{Pi-reduction}
    \fr\Pi^f{\bm\Omega}\cdot{\bm\Omega} 
    =\dot\theta^2\int_0^L\!\!\int_0^h \fr \rho (x^2+y^2)\rd y \rd x =
    \frac{1}{2}\left[\int_0^L \Big( x^2h + \mbox{$\frac{1}{3}$} h^3\Big)\,\rd x\right]\, \dot\theta^2 := \fr\mathbb{I}^f\dot\theta^2\,.
    \end{equation}
    
  The potential energy density for the fluid is
  \begin{equation}\label{PE-fluid-def}
    \begin{array}{rcl}
      \widehat{\textsf{PE}^f} &=& \rho g {\bf E}_2\cdot(\Rg{\bf x}+{\bf q})\\[2mm]
      &=& \rho g \Rg^T{\bf E}_2\cdot ({\bf x}+\Rg^T{\bf q})\\[2mm]
      &=& \rho  \Gamma\cdot ({\bf x}+{\bf r})\,,
    \end{array}
  \end{equation}
  and so
  \[
  \textsf{PE}^f =\int_0^L\!\!\int_0^h \rho  \Gamma\cdot ({\bf x}+{\bf r})\,
  \rd y\rd x = \Gamma\cdot\big(m_f\overline{\bf x}^f+m_f{\bf r}\big)\,.
  \]

  A similar construction gives the kinetic and potential energies of
    the vessel
    \[
    \begin{array}{rcl}
      \mathcal{L}^v &=& \textsf{KE}^v - \textsf{PE}^v \\[4mm]
      &=&\displaystyle \fr\Pi^v{\bm\Omega}\cdot{\bm\Omega} - m_v\overline{\bf x}^v\cdot
    {\bm\Omega}\times{\bf v} + \fr m_v \|{\bf v}\|^2 - m_v\Gamma\cdot(
                \overline{\bf x}^v + {\bf r})\,,
    \end{array}
    \]
                where $\rho_v$ is the density of the vessel material,
                \[
  m_v = \int_{V}\rho_v \,\rd x \rd y\,,\qand
  \overline{\bf x}^v =\frac{1}{m_v}\int_{V}\rho_v {\bf x}\,\rd x \rd y\,,
  \]
  where $\int_{V}(\cdot)\rd x\rd y$ is the integral over the vessel volume,
  and
  \[  
  \Pi^v{\bm\Omega}\cdot{\bm\Omega} := \left(\int_{V}\rho_v\Big( \boldsymbol{1}\|{\bf x}\|^2- {\bf x}{\bf x}^T\Big)\rd x \rd y\,{\bm\Omega}\right)\cdot{\bm\Omega} =
  \dot\theta^2\int_{V} \rho_v(x^2+y^2)\,\rd x\rd y\,.
  \]
              
  The full combined Lagrangian, $\mathcal{L}=\mathcal{L}^f+\mathcal{L}^v$, is
  \begin{equation}\label{L-psi-density}
    \begin{array}{rcl}
      \mathcal{L} &=&\displaystyle
      \int_0^L\int_0^h \Big( \fr \rho \|\nabla\psi\|^2 + \rho{\bf J}^T\nabla\psi\cdot{\bm\Omega}\times{\bf x} + \rho {\bf v}\cdot{\bf J}^T\nabla\psi  \Big)\rd y \rd x
    \\[6mm]
    &&\displaystyle\hspace{1.5cm}
    + \fr \big(\Pi^f+\Pi^v){\bm\Omega} \cdot{\bm\Omega}- (m_f\overline{\bf x}^f+m_v\overline{\bf x}^v\big)\cdot {\bm\Omega}\times{\bf v} + \fr (m_f+m_v) \|{\bf v}\|^2\\[4mm]
    &&\displaystyle\hspace{3.0cm}
    -  \Gamma\cdot\big( m_f\overline{\bf x}^f +m_v\overline{\bf x}^v+ (m_f+m_v){\bf r}\big) \,,
    \end{array}
  \end{equation}
 where the velocity field has been replaced by its stream function representation
  \begin{equation}\label{J_nabla}
  {\bf u} = {\bf J}^T\nabla\psi\,,\quad {\bf J}=\left[\begin{matrix} 0
      & -1 \\ 1 & 0 \end{matrix}\right]\,.
  \end{equation}
  In (\ref{L-psi-density}) $\psi$ is any smooth function on $\mathcal{D}$
  satisfying the boundary conditions
  \begin{equation}\label{psi-bc}
    \psi_y=0\ \mbox{at}\ x=0,L\,,\quad
    \psi_x=0\ \mbox{at}\ y=0\,,\qand \psi_x = -h_t-\psi_yh_x\ \mbox{at}\ y=h\,.
  \end{equation}
  These boundary conditions still leave the freedom to add an arbitrary
  function of time to $\psi$, and this value is fixed by taking $\psi=0$
  on the rigid boundaries
  \begin{equation}\label{psi-rigid-bc}
  \psi(x,0,t) = \psi(0,y,t) = \psi(L,y,t)=0 \,.
  \end{equation}
  
  When the fluid is neglected and $\Gamma$ is set to zero
  the Lagrangian density (\ref{L-psi-density}) reduces to the standard form
  for a Lagrangian that is left-invariant with respect to the special
  Euclidean group $SE(2)$, which generates the equations for rigid-body
  motion undergoing rotation and translation in the plane; see Chapter 7
  of \cite{holm-II} for the theory of $SE(n)-$invariant Lagrangians.

\section{Variations}
\setcounter{equation}{0}
\label{sec-variations}

The Lagrangian variables for the coupled problem are the fluid motion $\varphi$ as well as the position and orientation of the vessel $\mathbf{q}$ and $\mathcal{R}$. The variational principle in the Lagrangian path formulation is
\begin{equation}\label{HP_coupled}
\delta\int_{\tau_1}^{\tau_2} \widehat{\mathcal{L}}(\varphi, \dot\varphi, \mathcal{R},\dot{\mathcal{R}},\mathbf{q}, \dot{\mathbf{q}})\, \rd \tau = 0\,,
\end{equation}
with respect to \textit{free} variations $\delta\varphi$, $\delta\mathcal{R}$, $\delta\mathbf{q}$, vanishing at $\tau=\tau_1,\tau_2$, where $\tau=t$ in
the LPP setting. As indicated in the integrand of \eqref{HP_coupled}, to apply this principle, which is the natural extension of Hamilton's principle, one needs to express the Lagrangian \eqref{Lagr-density} in terms of the Lagrangian variables. The Lagrangian density
(\ref{L-psi-density}) is a reduced Eulerian form of (\ref{HP_coupled}) and forms
the basis for the variational principle here
\[
\delta\int_{t_1}^{t_2} \mathcal{L}(\psi,h,{\bm\Omega},{\bf r},{\bf v},{\bf \Gamma})\,\rd t = 0\,,
\]
with respect to appropriate \textit{constrained} variations $\delta \psi, \delta h, \delta{\bm\Omega}, \delta{\bf r}, \delta{\bf v}, \delta{\bf \Gamma}$.
Taking variations
\begin{equation}\label{delta-L-1}
        \begin{array}{rcl}
        0 &=&\displaystyle \int_{t_1}^{t_2}\left[
        \int_0^L\int_0^h\frac{\delta\mathcal{L}}{\delta\psi}\delta\psi\,\rd y
        \rd x +
        \int_0^L\frac{\delta\mathcal{L}}{\delta h}\delta h\,\rd x\right.\\[6mm]
          &&\left.\displaystyle\hspace{2.0cm}
        +\frac{\delta\mathcal{L}}{\delta{\bm\Omega}}\cdot\delta{\bm\Omega} +
        \frac{\delta\mathcal{L}}{\delta{\bf r}}\cdot\delta{\bf r}
        +\frac{\delta\mathcal{L}}{\delta{\bf v}}\cdot\delta{\bf v}
        +\frac{\delta\mathcal{L}}{\delta\Gamma}\cdot\delta\Gamma \right] \rd t\,.
        \end{array}
\end{equation}
However, setting each of these functional derivatives to zero does not recover
the governing equations, since the variations are not free.  In this section the
variations needed are recorded and they are justified from the Lagrangian particle path formulation \eqref{HP_coupled} in \S\ref{sec-reduction}.
The required fluid variations are
        \begin{equation}\label{fluid-variations}
        \begin{array}{rcl}
          \delta\psi &=&  w_t + \{w,\psi\}\quad\mbox{($0<y<h$, $0<x<L$)}\\[2mm]
          \delta\psi &=& W_t + \psi_y|^{y=h}\, W_x\quad\mbox{($y=h$)}\\[2mm]
          \delta\psi &=& 0\quad\mbox{($y=0$, $x=0,L$)}\\[2mm]
          \delta h &=& - W_x\,,
        \end{array}
        \end{equation}
        where $w(x,y,t)$ is a free variation in the interior with $w_y=0$ at $x=0,L$; $w=0$ at $y=0$; $w=0$ at $t=t_1,t_2$, and
        $W(x,t)=w(x,h(x,t),t)$ is the restriction
        of $w$ to the free surface. As it should, these variations are compatible with the free surface boundary condition (\ref{kfsbc}). Indeed, we have
          \[
          \begin{array}{rcl}
            0 &=& \delta(h_t+\Psi_x) = (\delta h)_t + (\delta\Psi)_x \\[2mm]
            &=& (\delta h)_t + (\delta\psi|^{y=h} + \psi_y|^{y=h}\delta h)_x \\[2mm]
&=& (-W_x)_t + (W_t + \psi|^{y=h}W_x - \psi_y|^{y=h}W_x)_x \\[2mm]
            &=& (-W_x)_t + (W_t)_x=0\,.
          \end{array}
          \]
          The vessel variations are
        \begin{equation}\label{vessel-variations}
        \begin{array}{rcl}
          \delta {\bm\Omega} &=& {\bm \Lambda}_t + {\bm\Omega}\times{\bm\Lambda}\\[2mm]
          \delta{\bf r} &=& {\bm\lambda} + {\bf r}\times{\bm \Lambda}\\[2mm]
          \delta{\bf v} &=& {\bm\lambda}_t + {\bm\Omega}\times{\bm\lambda} +
          {\bf v}\times{\bm \Lambda}\\[2mm]
            \delta{\bm\Gamma} &=& {\bm \Gamma}\times{\bm\Lambda}\,,
        \end{array}
        \end{equation}
          where ${\bm\Lambda}$ and ${\bm\lambda}$ depend on
          time only and are free variations vanishing at $t=t_1,t_2$.  Substitute these variations into
          (\ref{delta-L-1}) noting that ${\bm\Omega}\times{\bm\Lambda}=0$ in the 2D case,
\[
        \begin{array}{rcl}
        0 &=&\displaystyle\!
        \int_{t_1}^{t_2}\!\left[\int_0^L\!\int_0^h\frac{\delta\mathcal{L}}{\delta\psi}(w_t+\{w,\psi\})\,\rd y
        \rd x +
        \int_0^L\frac{\delta\mathcal{L}}{\delta\psi|^{y=h}}(W_t+uW_x)\,
        \rd x +
        \int_0^L\frac{\delta\mathcal{L}}{\delta h}(-W_x)\,\rd x\right.\\[6mm]
          &&\left.\quad \displaystyle
        +\frac{\delta\mathcal{L}}{\delta{\bm\Omega}}\cdot{\bm\Lambda}_t +
        \frac{\delta\mathcal{L}}{\delta{\bf r}}\cdot({\bm\lambda}+{\bf r}\times{\bm\Lambda})
        +\frac{\delta\mathcal{L}}{\delta{\bf v}}\cdot({\bm\lambda}_t + {\bm\Omega}\times{\bm\lambda} +
        {\bf v}\times{\bm\Lambda})
        +\frac{\delta\mathcal{L}}{\delta\Gamma}\cdot(\Gamma\times{\bm\Lambda})\right]\rd t\,.
        \end{array}
        \]
        Now include integration over $t$, integrate by parts, and use
        fixed endpoint conditions on the variations ${\bm\Lambda}$ and ${\bm\lambda}$.
        The abstract equations emerging are
        \[
        \begin{array}{rcl}
        &&\displaystyle  \delta w:\quad \frac{D\ }{Dt}\left( \frac{\delta\mathcal{L}}{\delta\psi}\right)=0\,,\\[6mm]
        &&\displaystyle  \delta W:\quad \frac{\partial\ }{\partial t}\left(
        \frac{\delta\mathcal{L}}{\delta\psi|^{y=h}}\right) +
        \frac{\partial\ }{\partial x}\left(u\frac{\delta\mathcal{L}}{\delta\psi|^{y=h}} - \frac{\delta\mathcal{L}}{\delta h}\right) = 0 \,,\quad \mbox{at}\ y=h\,,\\[6mm]
        &&\displaystyle \delta{\bm\Lambda}:\quad
        \frac{d }{dt}\left(
        \frac{\delta\mathcal{L}}{\delta{\bm\Omega}}\right)
        +{\bf r}\times \frac{\delta\mathcal{L}}{\delta{\bf r}}
        +{\bf v}\times \frac{\delta\mathcal{L}}{\delta{\bf v}}
        +{\bm\Gamma}\times \frac{\delta\mathcal{L}}{\delta{\bm\Gamma}}=0\,,\\[6mm]
        &&\displaystyle \delta{\bm\lambda}:\quad 
        \frac{d\ }{dt}\left( \frac{\delta\mathcal{L}}{\delta{\bf v}}\right) +{\bm\Omega}\times\frac{\delta\mathcal{L}}{\delta{\bf v}} =
          \frac{\delta\mathcal{L}}{\delta{\bf r}} \,.
          \end{array}
          \]
          The $\delta w$ equation gives the fluid equation in the interior and the $\delta W$ equation generates the GKK conservation law.
          The $\delta{\bm\Lambda}$ and $\delta{\bm\lambda}$ equations generate the
          rigid body motion of the vessel with fluid coupling.  The equations
          are accompanied by the boundary conditions (\ref{psi-bc}).
        
          The variational derivatives, obtained by differentiating
          (\ref{L-psi-density}), are
        \begin{equation}\label{delta-psi}
        \frac{\delta\mathcal{L}}{\delta\psi} = -\rho(\Delta\psi - 2\dot\theta)\,,
        \quad 0<y<h\,,
        \end{equation}
        in the interior, and at the free surface
        \begin{equation}\label{delta-psi-freesurface}
        \frac{\delta\mathcal{L}}{\delta\psi} = K\qand
        \frac{\delta\mathcal{L}}{\delta h} = \Xi\quad\mbox{at $y=h$}\,.
        \end{equation}
        A derivation of these variations starting with (\ref{L-psi-density})
        is given in Appendix \ref{app-a}.

         The variational derivatives associated with the rigid body motion are
        \begin{equation}\label{var-deriv-vessel}
        \begin{array}{rcl}
        &&\displaystyle \frac{\delta\mathcal{L}}{\delta{\bm\Omega}} =
        (\Pi^f+\Pi^v){\bm\Omega} + (m_f\overline{\bf x}^f+m_v\overline{\bf x}^v)\times{\bf v}+ \int_0^L\int_0^h ({\bf x}\times \rho{\bf u})\,\rd y\rd x\,,\\[6mm]
        &&\displaystyle \frac{\delta\mathcal{L}}{\delta{\bf r}} = -(m_f+m_v){\bm\Gamma}\,,\\[6mm]
        &&\displaystyle \frac{\delta\mathcal{L}}{\delta{\bf v}} =
          (m_f+m_v){\bf v} + {\bm\Omega}\times(m_f\overline{\bf x}^f+m_v\overline{\bf x}^v) + \int_0^L\int_0^h \rho{\bf u}\rd y \rd x\,,\\[6mm]
             &&\displaystyle \frac{\delta\mathcal{L}}{\delta{\bm\Gamma}} = -g
        (m_f\overline{\bf x}^f+m_v\overline{\bf x}^v+(m_f+m_v){\bf r})\,.
        \end{array}
        \end{equation}
        Substituting these expressions into the $\delta{\bm\Lambda}$
        and $\delta{\bm\lambda}$ equations gives the equations for the
        vessel coupled to the fluid motion, and they recover exactly
        the conservation of total linear and angular momentum in
        \S\ref{sec-goveqns}\ref{subsec-vessel-eqns}.
        Here the governing equation for ${\bf v}$ is expanded
        \begin{equation}\label{v-equation}
        \begin{array}{rcl}
          &&\displaystyle\frac{d\ }{dt}\left( (m_f+m_v){\bf v} + {\bm\Omega}\times(m_f\overline{\bf x}^f+m_v\overline{\bf x}^v) + \int_0^L\int_0^h \rho{\bf u}\rd y \rd x \right)
          \\[4mm]
          &&\displaystyle\hspace{1.0cm} + {\bm\Omega}\times\left( (m_f+m_v){\bf v} + {\bm\Omega}\times(m_f\overline{\bf x}^f+m_v\overline{\bf x}^v) + \int_0^L\int_0^h \rho{\bf u}\rd y \rd x\right) \\[4mm]
&&\hspace{2.0cm}          = -(m_f+m_v){\bm\Gamma}\,,
        \end{array}
        \end{equation}
        where ${\bf u}=(\psi_y,-\psi_x)$.        
        This expression generates only two component equations as the third
        component is identically zero.
        A similar expanded formula can be developed for the ${\bm\Omega}$ equation,
        with the ${\bm\Omega}$ equation having only one non-zero component.  When
        the velocity field ${\bf u}$ is restricted to be irrotational, the
        vessel equations agree with those in \cite{haa-thesis} and \cite{haa17}.

  \section{Special cases of governing equations}
  \setcounter{equation}{0}
  \label{sec-specialcases}

  When the vessel motion vanishes the Lagrangian (\ref{L-psi-density})
  reduces to
  \[
  \mathcal{L} =
  \int_0^L\int_0^h \Big( \fr \rho \|\nabla\psi\|^2 -\rho g y\Big)\rd y \rd x\,,
  \]
  with 
  \[
  \delta\mathcal{L} = \int_0^L\int_0^h \rho \nabla\psi\cdot\nabla\delta\psi \,\rd y \rd x + \int_0^L\Big[\fr\rho \|\nabla\psi\|^2 -\rho g y\Big]\Big|^{y=h}\delta h\,\rd x\,.
  \]
  Substituting for $\delta\psi$ and $\delta h$ from (\ref{fluid-variations}),
  and adding in the boundary conditions (\ref{psi-bc}),
  recovers the governing equations for the fluid in
  \S\ref{sec-goveqns}\ref{subsec-summary-fluideqns} with the
  GKK conservation law in the form (\ref{gkk-original}).

  When the
  fluid motion vanishes the vessel motion equations reduce to
         \begin{equation}\label{EP-vessel-eqns}
         \begin{array}{rcl}
         &&\displaystyle\Pi^v\bO_t +m_v\overline{\bf x}^v\times{\bf v}_t  + m_v({\bf v}\cdot\overline{\bf x}^v)\bO -
           m_v\bm\Gamma\times\overline{\bf x}^v=0\\[4mm]
           &&\displaystyle m_v{\bf v}_t + m_v\big(\bO\times{\bf v}\big)
           + m_v(\bO_t\times\overline{\bf x}^v)
        +m_v\bm\Gamma
        -m_v\bO\cdot\bO\overline{\bf x}^v =0\\[4mm]
        &&\bm\Gamma_t +\bO\times\bm\Gamma = 0\,.
         \end{array}
         \end{equation}
         These equations are similar to Kirchoff's equations for a
         rigid body in moving in 3D (see Chapter
         6 of \textsc{Lamb}~\cite{lamb} and \S7.2 in \textsc{Holm}~\cite{holm-II}).  The latter
         view of the equations (\ref{EP-vessel-eqns}) is that they are
         the Euler-Poincar\'e equations for a Lagrangian that is
         left-invariant with respect to the group $SE(2)$, the special
         Euclidean group in the plane, although here there is the additional term due
         to gravity and represented by the vector ${\bm\Gamma}$.
         \begin{figure}[!ht]
         \centering\includegraphics[width=5.5cm]{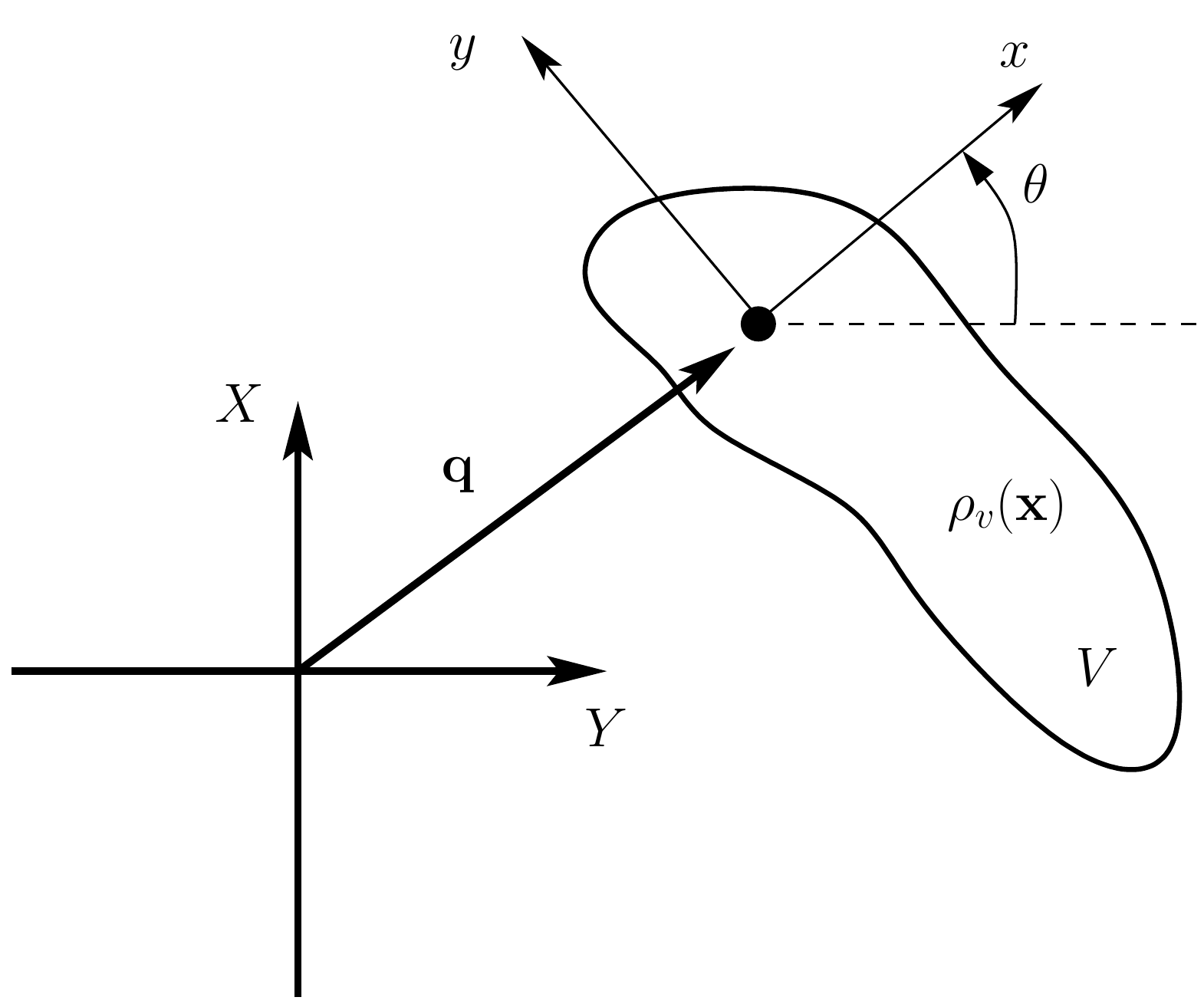}
           \caption{Schematic of rigid body motion in a vertical plane.}
         \label{fig-2d-rigidbody}
\end{figure}
         In 2D these equations are just the
         governing equations of a compound pendulum in the plane with
         translating pivot point (see Figure \ref{fig-2d-rigidbody}), or more generally a rigid body moving in
         the plane subject to a gravitational field. In 2D, the first equation
         of (\ref{EP-vessel-eqns}) has only one nonzero component and the
         second has only two nonzero components.

\section{Justifying constrained variations via reduction}
\setcounter{equation}{0}
\label{sec-reduction}

The natural variational principle in fluid mechanics is
Hamilton's principle in the LPP setting.  Natural because the density
is the kinetic minus potential energy and the variations are free.
In the LPP setting the position of particles is determined by a
mapping $\varphi :B\to \mathcal{D}$ where $B$ is a reference space
\[
B = \{{\bf a}=(a_1,a_2)\in \R^2\ :\ 0\leq a_1\leq 1\,,\ 0\leq a_2\leq 1\}
\]
and $\mathcal{D}$ is the current fluid configuration (\ref{fluid domain}).
The position and velocity of a fluid particle at any time $\tau$ in
the LPP setting are 
\begin{equation}\label{lpp-x}
  {\bf x} = \varphi({\bf a},\tau)\qand \mathbf{u}^{\textsf{Lag}}({\bf a},\tau) = \dot\varphi({\bf a},\tau)\,.
  \end{equation}
$\tau=t$ but the distinction is maintained as $\partial_\tau$ is
taken with $\mathbf{a}$ fixed and $\partial_t$ is taken with $\mathbf{x}$
fixed.

The LPP Lagrangian variational principle for the fluid only is then
\begin{equation}\label{HP}
\delta\int_{\tau_1}^{\tau_2} \mathcal{L}(\varphi, \dot{\varphi}) \rd \tau=0,
\end{equation}
with respect to arbitrary variations $\delta \varphi$ vanishing at $\tau=\tau_1,\tau_2$.  To be precise the configuration  manifold is an appropriate subset of
the manifold of volume-preserving embeddings (e.g.\ \cite{cl89,gbms12,gbv15} and references therein) but this level of detail will not
be needed here.

The free surface is the image of the upper edge of the reference space
\begin{equation}\label{Sigma-def}
\Sigma^s = \{ (x,y)\in\R^2 \ :\ x = \varphi_1(a_1,1,\tau)\ \mbox{and}\ y=\varphi_2(a_1,1,\tau)\}\subset\partial\mathcal{D}\,.
\end{equation}

Variation of $\mathcal{L}$ in (\ref{L-psi-density}) will require variations in the Eulerian setting.
The transformation from the LPP description
to the Eulerian description is a form of reduction, obtained by factoring
out the particle relabelling group (cf. \S11.1 of \cite{hss09} and
\cite{gbms12} when a free boundary is included).  This
reduction converts free variations in the LPP setting to constrained
variations in the Eulerian setting.

The Lagrangian and Eulerian velocity fields are related by
\begin{equation}\label{uL-uE-relation}
\mathbf{u}(\mathbf{x},t) = \dot\varphi ( \varphi^{-1}(\mathbf{x},t),t)
  \quad\Rightarrow\quad
  \dot\varphi({\bf a},\tau) = \mathbf{u}(\varphi({\bf a},\tau),\tau)\,,
\end{equation}
and $\mathbf{u}$ is a divergence-free vector field on $\mathcal{D}$,
parallel to rigid boundaries but not parallel to $\Sigma^s$.
Let $\delta\varphi ({\bf a},\tau)$ be a free variation of the fluid displacement
in the LPP setting
and let ${\bf z}(\mathbf{x},t)$ be its representation in
the Eulerian setting,
\begin{equation}\label{z-varphi-def}
  {\bf z}(\mathbf{x},t) =\delta\varphi(\varphi^{-1}(\mathbf{x},t),t)\quad\Rightarrow\quad
  \delta \varphi({\bf a},\tau)= \mathbf{z}(\varphi({\bf a},\tau),\tau)\,.
\end{equation}
The field $\mathbf{z}$ is an arbitrary divergence-free
vector field on $\mathcal{D}$, parallel to rigid boundaries
but not parallel to $\Sigma^s$.

To derive (\ref{delta-u-def-intro}) vary (\ref{uL-uE-relation}),
  \begin{equation}\label{delta-varphi}
    \delta\dot\varphi(\mathbf{a},\tau) =
    \delta\mathbf{u}(\varphi(\mathbf{a},\tau),\tau) + \delta\varphi(\mathbf{a},\tau)\cdot\nabla{\bf u}(\varphi(\mathbf{a},\tau),\tau)\,.
  \end{equation}
  Now differentiating (\ref{z-varphi-def}) with respect to $t$ gives
  \begin{equation}\label{dot-varphi}
    \delta\dot\varphi(\mathbf{a},\tau) = {\bf z}_t(\varphi(\mathbf{a},\tau),\tau)
    +\dot\varphi(\mathbf{a},\tau)\cdot\nabla\mathbf{z}(\varphi(\mathbf{a},\tau),\tau)\,.
  \end{equation}
  Combining these two expressions then gives
\begin{equation}\label{delta-u-def}
\delta\mathbf{u} = {\bf z}_t + [\mathbf{u},\mathbf{z}]\quad
\mbox{with}\quad
  [\mathbf{u},\mathbf{z}] := \mathbf{u}\cdot\nabla\mathbf{z}-
  \mathbf{z}\cdot\nabla\mathbf{u}\,.
\end{equation}
In forming this equation, ${\bf a}$ is replaced by ${\bf a}= \varphi^{-1}(\mathbf{x},t)$
rendering it a purely Eulerian expression.
The identity (\ref{delta-u-def}) is one of the most important, but unheralded,
identities in fluid mechanics as it relates small changes in the Eulerian
velocity field to small changes in the Lagrangian velocity field, all
viewed from the Eulerian perspective.
Further detail on this identity, its history, abstraction,
and generalisation can be found in \cite{hmr98,hss09} and references
therein.

The two new results on Eulerian variations needed in this paper are
the implications for (\ref{delta-u-def}) when the velocity field is
represented by a stream function, and the induced free surface
variation when the surface $\Sigma^s$ is represented by a graph.

\subsection{Stream function variations}
\label{subsec-streamf-variations}

The aim is to reduce
\begin{equation}\label{eqn6.1}
\delta\mathbf{u} = {\bf z}_t + [\mathbf{u},\mathbf{z}]\quad
\mbox{to}\quad \delta \psi=w_t+\{w,z\}\,.
\end{equation}
Define $w(x,y,t)$ by expressing
the divergence-free vector field $\mathbf{z}$ in terms of a $w$
stream function
\begin{equation}\label{z-w-def}
\mathbf{z}:=\mathbf{J}^T\nabla w\,.
\end{equation}
The properties of ${\bf z}$ then give that
$w$ is an arbitrary scalar-valued function on $\mathcal{D}$
satisfying $w_y=0$ at $x=0,L$; $w_x=0$ at $y=0$; and $w=0$ at $t=t_1,t_2$.

The $w$ stream function is obtained from a given
$\mathbf{z}$ by integrating
\[
dw = -z_2\rd x + z_1\rd y\,,
\]
along a curve in $\mathcal{D}$. By choosing a reference value of
$w$, e.g.\ fixing $w$ on rigid boundaries,
\begin{equation}\label{w-rigid-bc}
  w(x,0,t)=w(0,y,t)=w(L,y,t)=0\,,
\end{equation}
the $w$ stream function is unique. This condition mirrors the condition on $\psi$ in (\ref{psi-rigid-bc}).
  
Inserting the velocity representation $\mathbf{u}=\mathbf{J}^T\nabla\psi$ and
(\ref{z-w-def}) into the first expression in (\ref{eqn6.1}) gives
\begin{equation}\label{delta_J}
\mathbf{J}^T\nabla \delta \psi= \mathbf{J}^T\nabla w_t+[\mathbf{J}^T\nabla\psi,\mathbf{J}^T\nabla w].
\end{equation}
A direct computation then shows that $[\mathbf{J}^T\nabla\psi,\mathbf{J}^T\nabla w]= \mathbf{J}^T\nabla\{w,\psi\}$, and so \eqref{delta_J} yields
\[
\mathbf{J}^T\nabla (\delta \psi- w_t-\{w,\psi\})=0\,,
\]
or
\begin{equation}\label{delta-psi-f}
\delta \psi = w_t+\{w,\psi\} + f(t)\,,
\end{equation}
where $f(t)$ is in general arbitrary.
However, evaluating (\ref{delta-psi-f}) at $y=0$ and using
the normalization (\ref{w-rigid-bc}) gives
$f(t)$=0.  This confirms the form for $\delta\psi$
in the second expression in (\ref{eqn6.1}).

\subsection{Free surface variations}
\label{subsec-freesurface}

        With the free surface represented by a graph, $y=h(x,t)$, the
        mapping from LPP to Eulerian variables (\ref{Sigma-def}) on
        $\Sigma^s$ becomes
        \begin{equation}\label{def_h}
        h(\varphi_1(a_1,1,\tau),\tau) = \varphi_2(a_1,1,\tau)\,.
        \end{equation}
        The time derivative of this equality gives the free surface condition \eqref{fsbcs}.
        Taking variations of \eqref{def_h}, we get
        \begin{equation}\label{delta-h-lpp}
        \delta h(\varphi_1(a_1,1,\tau),\tau)+
        h_x \delta\varphi_1(a_1,1,\tau) = \delta \varphi_2(a_1,1,\tau)\,.
        \end{equation}
        But using the stream function representation for ${\bf z}$, we get
        \[
        \delta\varphi({\bf a},\tau) = \mathbf{z}(\varphi({\bf a},\tau),\tau)
        =(w_y(\varphi(a_1,a_2,\tau),\tau),-w_x(\varphi(a_1,a_2,\tau),\tau))
        \]
        and hence
        \[
        \delta\varphi_1(a_1,1,\tau) = w_y(\varphi(a_1,1,\tau),\tau)
        \qand
        \delta\varphi_2(a_1,1,\tau) = -w_x(\varphi(a_1,1,\tau),\tau)\,,
        \]
        so the second and third terms in (\ref{delta-h-lpp}) combine into
        \[
        h_x\delta\varphi_1(a_1,1,\tau)-\delta\varphi_2(a_2,1,\tau) =
        h_x w_y(\varphi_1(a_1,1,\tau),1,\tau) + w_x(\varphi_1(a_1,1,\tau),1,\tau)\,.
        \]
        In Eulerian variables the right-hand side is
        \[
        \Big[h_x w_y(\varphi_1,1,\tau) + w_x(\varphi_1,1,\tau)\Big]\Big|^{a_2=1}
        \xrightarrow{\varphi^{-1}} h_x w_y(x,h(x,t),t) + w_x(x,h(x,t),t) = W_x\,.
          \]
          Substution into (\ref{delta-h-lpp}) and mapping $\delta h$ to
          Eulerian variables then gives
          \[
          \delta h = - W_x\,,
          \]
          confirming the expression in (\ref{fluid-variations}).

          \subsection{Variation of vessel parameters}

          The variations of ${\bm\Omega}$, ${\bf r}$, ${\bf v}$, and ${\bm\Gamma}$
          in (\ref{vessel-variations})
          arise in rigid body motion and the details can be found in
          \cite{hss09}.  Here, just a sketch of the basic idea is given.
          A detailed derivation of $\delta{\bm\Omega}$ is given on pages 249-250
          of \cite{hss09} with $\widehat{\bm\Lambda} :=\Rg^{T}\delta\Rg$
          where $\widehat{\bm\Lambda}$ is the representation of the vector $\bm\Lambda$ in
          terms of a $3\times 3$ skew-symmetric matrix.  Let
          ${\bm\lambda}=\Rg^T\delta{\bf q}$ then
        \[
        \delta{\bf r} = \delta\big( \Rg^T{\bf q}\big) =
         -\Rg^T\delta\Rg\Rg^T{\bf q} +\Rg^T\delta {\bf q}
         = {\bm\lambda} - \widehat\Lambda{\bf r}
         = {\bm\lambda} +{\bf r}\times{\bm\Lambda}\,.
         \]
         A similar argument gives the expression for $\delta{\bf v}$
         in (\ref{vessel-variations}).  For the gravity vector
         \[
         \delta{\bm \Gamma }= \delta\big( \Rg^T{\bf e}_2\big) =
         -\Rg^T\delta\Rg\Rg^T{\bf e}_2 = -\Rg^T\delta\Rg{\bm\Gamma}=
         -{\bm\Lambda}\times{\bm\Gamma}\,,
         \]
         confirming the fourth equation in (\ref{vessel-variations}).

  \section{Concluding remarks}
  \setcounter{equation}{0}
  \label{sec-cr}

  In this paper a new variational principle has been introduced for 2D
  inviscid incompressible fluid flow in a moving
  vessel with coupling to the
  vessel motion.  The variational principle is useful for identifying
  conservation laws, devising approximate schemes, and contributes to the
  design of numerical methods.

\begin{figure}[!h]
\centering\includegraphics[width=5cm]{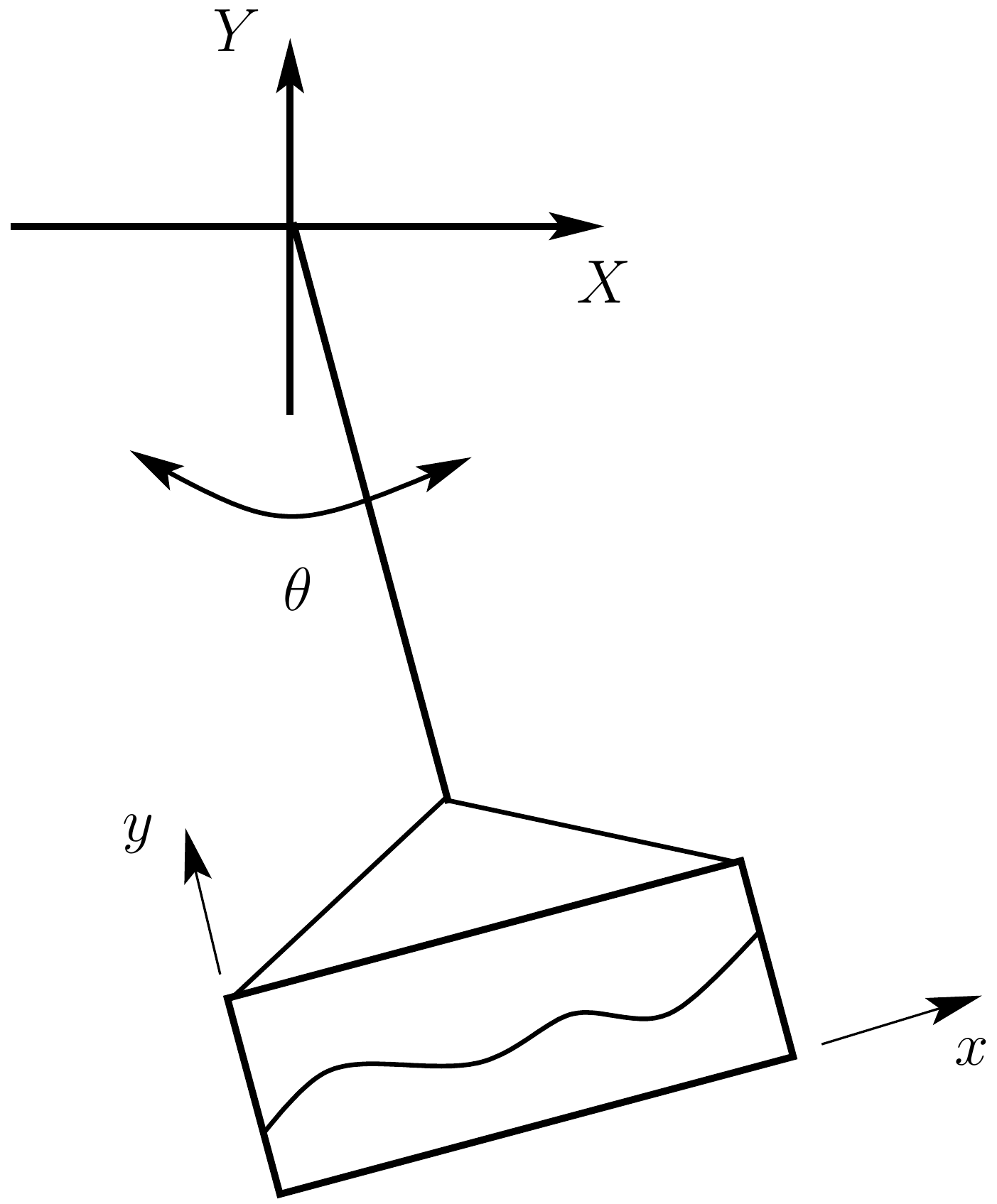}
\caption{The pendulum slosh problem with vorticity.}
\label{fig-pendulum-slosh}
\end{figure}

  An open question is to identify canonical problems where vorticity is
         important in sloshing.  \textsc{Timokha}~\cite{timokha} discusses
         vorticity in sloshing, and discusses the ``glass-wine'' paradox,
         whereby a steady-state swirl motion in a vessel generates a vortex
  by conversion of the wave angular momentum to the vortex angular
  momentum.  This latter motion is in 3D.
  In 2D vorticity is observed in experiments when baffles are
  introduced, and this could be modelled by appropriate introduction
  of vorticity in the initial data.  A canonical test problem is the
  ``pendulum-slosh'' problem, illustrated in Figure \ref{fig-pendulum-slosh},
  where ${\bf q}=0$ and the only vessel degree of freedom is rotation.
  The irrotational case has been studied in \textsc{Turner et al.}~\cite{tbaa15} and the implication of adding vorticity to the initial data is of interest.

\vspace{-.25cm}
  The Lagrangian variational principle here opens the door to generating a
  Hamiltonian formulation with a Lie-Poisson structure, extending the theory
  in \textsc{Lewis et al.}~\cite{lmmr86} (for incompressible free
  surface flows) and
  \textsc{Mazer \& Ratiu}~\cite{mr89} (for compressible
  free surface flows) to the dynamically coupled problem.
  
  In the formulation presented here there
  were hints about the structure of the problem in 3D. However there
  are still technical difficulties in extending the fluid part
  to 3D.  Firstly, a
  parameterisation of divergence-free vector fields 
  is required.  The preferred option is a vector stream function, but the pressure
  boundary condition then presents challenges.  The tangential derivative of
  pressure has two components and the extension of the GKK theory to 3D does not
  result in a conservation law (see \cite{gkk15} for
  the GKK theory in 3D without rotation).
  On the other hand, the extension of the vessel motion
  to 3D is straightforward.

  \begin{appendix}

     \section*{Appendix A: Details of the $\delta\mathcal{L}$ calculations}
     \label{app-a}
     
     First look at the variations associated with the stream function with
     the other variables fixed.
    The stream function terms in $\mathcal{L}$ are
         \[
      \mathcal{L}^\psi =
      \int_0^L\int_0^h \Big( \fr \rho (\psi_x^2+\psi_y^2) -\rho\dot\theta(x\psi_x+y\psi_y)
      + \rho (v_1\psi_y-v_2\psi_x) \Big)\rd y \rd x\,.    
    \]
    Vary $\mathcal{L}^\psi$ with respect to $\psi$ keeping all other variables fixed, integrate by parts, and use $\delta\psi=0$ on rigid boundaries
    from (\ref{psi-bc}), then
    \[
    \begin{array}{rcl}
      \delta\mathcal{L}^\psi &=&\displaystyle
            -\int_0^L\int_0^h \rho (\Delta\psi-2\dot\theta)\delta\psi\,\rd y\rd x            +\int_0^L\rho (\psi_y-\dot\theta h + v_1)\delta\psi\bigg|^{y=h}\,\rd x\\[6mm]
            &&\displaystyle\hspace{2.0cm}
            +\int_0^L \rho h_x (-\psi_x+\dot\theta x + v_2)\delta\psi\bigg|^{y=h}\rd x\\[6mm]
            &=&\displaystyle
            -\int_0^L\int_0^h \rho (\Delta\psi-2\dot\theta)\delta\psi\,\rd y\rd x
            + \int_0^L K\,\delta\psi\big|^{y=h}\,\rd x\,.      
    \end{array}
    \]
    This confirms the variational derivatives (\ref{delta-psi}) and the
    first of (\ref{delta-psi-freesurface}).
    
 Now keep all other variations fixed, and vary with with respect to $h$,
          \[
          \begin{array}{rcl}
      \delta\mathcal{L} &=&\displaystyle
      \int_0^L\Big( \fr \rho \|{\bf u}\|^2 + \rho{\bf u}\cdot{\bm\Omega}\times{\bf x} + \rho {\bf v}\cdot{\bf u}  \Big)\delta h  + \fr \rho(x^2+h^2){\bm\Omega} \cdot{\bm\Omega}\delta h
    \\[6mm]
    &&\displaystyle\hspace{1.5cm}
    - \rho\big({\bf x}\big|^{y=h}\big)\cdot {\bm\Omega}\times{\bf v}\delta h + \fr \rho\delta h \|{\bf v}\|^2\delta h
    - \rho  \Gamma\cdot\big( {\bf x}\big|^{y=h} + {\bf r}\big)\delta h\,\rd x \,,
    \end{array}
    \]
    using
    \[
    \delta m_f = \int_0^L\rho \delta h\,\rd x\,,\quad \delta(m_f\overline{\bf x}^f) =
    \int_0^L\rho{\bf x}\big|^{y=h}\delta h\,\rd x
      \qand
      \delta \int_0^h\fr\rho(x^2+y^2)\,\rd y = \fr\rho(x^2+h^2)\delta h\,.
      \]
      Rearranging and substituting the above expressions
          \begin{equation}\label{deltaL-deltah}
          \frac{\delta\mathcal{L}}{\delta h} = \fr\rho(u-\dot\theta h+v_1)^2+
          \fr\rho(v+\dot\theta x + v_2)^2 -\rho  (\Gamma_1(x+r_1) + \Gamma_2(h+r_2))\,,
          \end{equation}
          which confirms the second of (\ref{delta-psi-freesurface}) using
          (\ref{Xi-condensed}).  
        
          For the variations associated with vessel motion, keep $\psi$ and $h$
          fixed and vary with respect to ${\bm\Omega}$, ${\bf r}$,
          ${\bf v}$ and $\bm\Gamma$,
 \[
    \begin{array}{rcl}
      \delta\mathcal{L} &=&\displaystyle
      \int_0^L\int_0^h \Big(\rho{\bf u}\cdot\delta{\bm\Omega}\times{\bf x} + \rho\, \delta{\bf v}\cdot{\bf u}  \Big)\rd y \rd x
      + \big(\Pi^f+\Pi^v){\bm\Omega} \cdot\delta{\bm\Omega}
    \\[6mm]
    &&\displaystyle\hspace{1.5cm}   - (m_f\overline{\bf x}^f+m_v\overline{\bf x}^v\big)\cdot \big( \delta{\bm\Omega}\times{\bf v} + {\bm\Omega}\times\delta{\bf v} \big)
    + (m_f+m_v) {\bf v}\cdot\delta{\bf v}\\[4mm]
    &&\displaystyle\hspace{3.0cm}
    -   (m_f+m_v) \bm\Gamma\cdot\delta{\bf r}
    -  \delta\bm\Gamma\cdot\big( m_f\overline{\bf x}^f +m_v\overline{\bf x}^v+ (m_f+m_v){\bf r}\big)\,.
    \end{array}
    \]
    Rearrange and group terms
    \[
    \begin{array}{rcl}
      \delta\mathcal{L} &=&\displaystyle
      \delta{\bm\Omega}\cdot\int_0^L\int_0^h (\rho{\bf x}\times{\bf u})\,\rd y\rd x
      +\delta{\bm\Omega}\cdot(m_f\overline{\bf x}^f+m_v\overline{\bf x}^v\big)\times{\bf v}
      + \big(\Pi^f+\Pi^v){\bm\Omega} \cdot\delta{\bm\Omega}
    \\[6mm]
    &&\displaystyle\hspace{1.5cm}
    +\delta{\bf v}\cdot\int_0^L\int_0^h\rho{\bf u}\,\rd y \rd x
    +{\bm\Omega}\times\big(m_f\overline{\bf x}^f+m_v\overline{\bf x}^v\big)\cdot\delta{\bf v}     + (m_f+m_v) {\bf v}\cdot\delta{\bf v}\\[6mm]
    &&\displaystyle\hspace{3.0cm}
    -   (m_f+m_v) \bm\Gamma\cdot\delta{\bf r}
    -  \delta\bm\Gamma\cdot\big( m_f\overline{\bf x}^f +m_v\overline{\bf x}^v+ (m_f+m_v){\bf r}\big)\,.
    \end{array}
    \]
    Extracting the variational derivatives,
    \[
    \delta\mathcal{L} = \frac{\delta\mathcal{L}}{\delta{\bm\Omega}}\cdot\delta{\bm\Omega}
+ \frac{\delta\mathcal{L}}{\delta{\bf v}}\cdot\delta{\bf v}
+ \frac{\delta\mathcal{L}}{\delta{\bf r}}\cdot\delta{\bf r}
+ \frac{\delta\mathcal{L}}{\delta\bm\Gamma}\cdot\delta\bm\Gamma\,,
\]
    confirms the formulae in (\ref{var-deriv-vessel}).

  \end{appendix}


\end{document}